\documentclass[prb,twocolumn,floatfix]{revtex4}
\usepackage{graphicx}

\begin{document}
\title{%
Hund's coupling and the metal-insulator transition in the two-band
Hubbard model
}
\author{Th.\ Pruschke}
\affiliation{Institute for Theoretical Physics, University of G\"ottingen, 
Friedrich-Hund-Platz 1,
37077 G\"ottingen, Germany}
\author{R.\ Bulla}
\affiliation{Center for electronic correlations and magnetism,
Theoretical Physics III, Institute of Physics, University of Augsburg, 
86135 Augsburg, Germany}

\begin{abstract}
The Mott-Hubbard metal-insulator transition is investigated in a two-band
Hubbard model within dynamical mean-field theory. To this end, we use a
suitable extension of Wilson's numerical renormalization group for the
solution of the effective two-band single-impurity Anderson model. This
method is non-perturbative and, in particular, allows to take into
account the full exchange part of the Hund's rule coupling between the
two orbitals. We discuss in detail the influence of the various
Coulomb interactions on thermodynamic and dynamic properties, for both
the impurity and the lattice model. The exchange part of the 
Hund's rule coupling turns out to play an important role for the physics
of the two-band Hubbard model and for the nature of the Mott-transition.
\end{abstract}
\pacs{}
\maketitle              

\section{Introduction}
Many materials with open $d$- or $f$-shells show metal-insulator transitions
which are commonly classified to be of Mott-Hubbard type due to strong 
electron-electron correlations \cite{imada98}.
The conventional model to study this type of transition is
the one-band Hubbard model \cite{hubbard}, which in standard notation reads
\begin{equation}
\label{equ:hubbard}
H=-\sum_{ij\sigma}t_{ij}
c^{\dagger}_{i\sigma}c^{\phantom{\dagger}}_{j\sigma}
+\frac{U}{2}\sum_{i\sigma}n_{i\sigma}n_{i\bar{\sigma}}\;\;.
\end{equation}
Major progress in understanding the physics of the Mott-Hubbard metal-insulator
transition (MHMIT) of the model (\ref{equ:hubbard}) has been achieved in the last decade through
the development of the dynamical mean-field theory (DMFT) \cite{mv,pradv,rmp}.
At $T\!=\!0$ the MHMIT occurs at a  critical value of the
Coulomb parameter $U_c\approx 1.5W$ \cite{jazph,rmp,bulprl}, where $W$
denotes the bandwidth of the density of states at $U=0$. Interestingly, the
transition is of first order \cite{rmp,bulcosvol} for $T>0$ with a second order
end point at a $T_c\approx 0.017W$ and $U_c\approx 1.2W$. 

Such a second order end point is also seen in the phase diagram of 
typical Mott-Hubbard systems like  V$_2$O$_3$ \cite{mcwhan}. Therefore,
the one-band Hubbard model has been frequently used as a microscopic
model for these materials (see Refs.~\onlinecite{rmp,roz95} and,
in particular, Ref.~\onlinecite{Limelette} 
which focusses on the critical regime close
to the second order end point). On the other hand, the justification
to base the microscopic description on a one-band model 
(see Ref.~\onlinecite{castellani}) has been questioned 
recently\cite{evshov}.

For a proper description of materials such as transition metal oxides,
the orbital structure of the relevant electronic degrees of freedom
has to be taken into account. This can lead to a fairly complicated
form of the underlying tight-binding bandstructure (the kinetic energy term
acquires a matrix structure). Furthermore, additional local Coulomb matrix 
elements arise which describe the interactions
between electrons in different orbitals.

The simplest possible extension of the model (\ref{equ:hubbard}) to the case
of orbital degeneracy is the two-band Hubbard model. It is a relevant
model whenever the electronic degrees of freedom close to the Fermi level
are two-fold degenerate, as for the $e_g^\sigma$ states in materials
like LaMnO$_3$ or KCuF$_3$ \cite{imada98}. Here we investigate the
two-band Hubbard model in the following form:
\begin{equation}
\label{equ:tbhubbard}
H=\begin{array}[t]{l}
\displaystyle-\sum_{ij}\sum_{mm'\sigma}t_{ij}^{mm'}
c^{\dagger}_{im\sigma}c^{\phantom{\dagger}}_{jm'\sigma}\\[5mm]
\displaystyle+\frac{U}{2}\sum_{i}\sum_{m\sigma}n_{im\sigma}n_{im\bar{\sigma}}\\[5mm]
\displaystyle+\frac{2U'-J}{4}\sum_{i}\sum_{m\ne m'}\sum_{\sigma\sigma'}n_{im\sigma}n_{im'\sigma'}\\[5mm]
\displaystyle-J\sum_{i}\sum_{m\ne m'}\vec S_m\cdot\vec S_{m'}\\[5mm]
\displaystyle-\frac{J}{2}
\sum_{i}\sum_{m\ne m'}\sum_\sigma c_{im\sigma}^\dagger
c_{im\bar\sigma}^\dagger 
c_{im'\bar\sigma}^{\phantom{\dagger}}
c_{im'\sigma}^{\phantom{\dagger}} \ \ ,
\end{array}\;\;
\end{equation}
with the orbital index $m=1,2$.
The Coulomb parameters $U'\ge0$ and $J\ge0$ describe the inter-orbital Coulomb
interaction and Hund's exchange coupling, respectively. The last term 
in (\ref{equ:tbhubbard}) 
is necessary to ensure rotational invariance of the interaction.
By virtue of this rotational invariance the Coulomb parameters are 
related by $U'=U-2J$. Generally, the hierarchy of interactions is
$U>U'>J$.

A multi-band Hubbard model as in eq.~(\ref{equ:tbhubbard}) displays
a Mott transition at all integer fillings (not only at half filling as
in the single-band case). The additional Coulomb interactions
also modify the value of $U_c$ and, possibly,
the character of the transition, as has already been investigated
within the DMFT framework \cite{roz97,han,kawa,florens,ohno}.

Various theoretical and numerical techniques have been employed
to investigate multi-band Hubbard models within DMFT. 
The Quantum Monte Carlo method \cite{roz97,han},
which is very successful in treating the multi-band Hubbard models
in present applications of the LDA+DMFT approach \cite{ldadmft}, 
cannot, however,
handle the rotationally invariant form of the interaction in 
(\ref{equ:tbhubbard}) due to the sign problem (for recent
progress in reducing this sign problem, 
see Ref.~\onlinecite{sign,lich}).
On the other hand, exact diagonalization \cite{kawa}, linearized
DMFT \cite{ohno} and exact treatments in the limit of infinite orbital 
degeneracy \cite{florens} are not able to reliably calculate dynamical
properties.

In this paper, we use  Wilson's numerical renormalization group
(NRG) \cite{nrg} to solve the effective two-orbital quantum impurity
problem which appears in the DMFT for the two-band Hubbard model.
In Sec.~II we start with some technical issues related to the
NRG in the two-band case. Section III shows thermodynamic
and dynamic quantities for the
two-orbital single-impurity Anderson model, with the focus
on the role of the Hund's coupling.
In Sec.~IV we discuss the two-band Hubbard model;
here we concentrate on the simplest case, i.e.\ degenerate orbitals
and intra-orbital hopping only:
$t_{ij}^{mm'}=t_{ij}\delta_{mm'}$. 
The model eq.~(\ref{equ:tbhubbard}) is studied on a Bethe lattice
(mainly to compare our results with those from other approaches);
the generalization to other lattices (other densities of states)
is straightforward. The paper is summarized in Sec.~V.

\section{NRG for multi-orbital models}

The use of the NRG to solve the effective quantum impurity model appearing 
in the DMFT self-consistency has been 
extensively discussed in the 
literature \cite{bulprhew,bulprl,bulcosvol}. Here we focus
on the additional problems arising in the two-orbital impurity Anderson model,
which constitutes the effective local model arising in the DMFT for the
two-orbital Hubbard model.

The two-orbital Anderson impurity Hamiltonian 
(in standard notation) is given by
\begin{equation}
\label{equ:tbsiam}
H=\begin{array}[t]{l}
\displaystyle\sum\limits_{\vec km\sigma}\epsilon_{\vec km\sigma}
c^{\dagger}_{\vec km\sigma}c^{\phantom{\dagger}}_{\vec km\sigma}
+\sum_{m\sigma} \epsilon_d
d^{\dagger}_{m\sigma}d^{\phantom{\dagger}}_{m\sigma}\\[5mm]
\displaystyle+\frac{U}{2}\sum_{m\sigma}n_{m\sigma}^dn_{m\bar{\sigma}}^d\\[5mm]
\displaystyle+\frac{2U'-J}{4}\sum_{m\ne m'}\sum_{\sigma\sigma'}n_{m\sigma}^dn_{m'\sigma'}^d\\[5mm]
\displaystyle-J\sum_{m\ne m'}\vec S_m\cdot\vec S_{m'}\\[5mm]
\displaystyle-\frac{J}{2}
\sum_{m\ne m'}\sum_\sigma d_{m\sigma}^\dagger
d_{m\bar\sigma}^\dagger 
d_{m'\bar\sigma}^{\phantom{\dagger}}
d_{m'\sigma}^{\phantom{\dagger}}\\[5mm]
\displaystyle\frac{V}{\sqrt{N}}\sum_{\vec km\sigma}
c^{\dagger}_{\vec km\sigma}d^{\phantom{\dagger}}_{m\sigma}+\mbox{h.c.}
\end{array}
\end{equation}

Within the NRG approach, the quantum impurity problem is
mapped onto a semi-infinite chain form \cite{nrg} with the impurity
at the first site of the chain and the conduction band
written in a one-dimensional tight-binding form. The model
in the semi-infinite chain form is then solved by iterative
diagonalization. Therefore, the major obstacle in applying the
NRG to multi-band models is the dramatic increase of the Hilbert
space with each NRG step (in each step, one additional site of the
semi-infinite chain is included).

A possible solution to this problem is to use very large values
of the discretization parameter $\Lambda$ so that the number of
states can be reduced significantly. Averaging over many discretizations
(the so-called ``Z-trick'') has been shown to give reliable results
for thermodynamic quantities (see Ref.~\onlinecite{oliveira}), even
for large values of $\Lambda$. However, this approach becomes
at least cumbersome for the calculation of dynamic quantities which are
required for the DMFT self-consistency.

Here we adopt two different strategies which allow to use a small
value of the NRG discretization parameter $\Lambda$ and
to keep enough states in each step so that dynamic quantities
can be calculated reliably. The first one is to explicitely
include the orbital quantum number in the iterative construction
of the basis states. This additional quantum number significantly
reduces the typical matrix size, so that O$(7000)$ states
can be kept in each NRG iteration with reasonable computation time and
memory consumption on modern SMP high-performance computers like e.g.\ the
IBM Regatta. The price to pay is that one has to omit the last term in
the Coulomb interaction, i.e.
$$
\frac{J}{2}
\sum_{m\ne m'}\sum_\sigma d_{m\sigma}^\dagger
d_{m\bar\sigma}^\dagger 
d_{m'\bar\sigma}^{\phantom{\dagger}}
d_{m'\sigma}^{\phantom{\dagger}}\;\;,
$$
because it explicitely breaks the orbital symmetry. However, it turns out
that this term does not influence the thermodynamics, and dynamic
quantities are only slightly affected via the multiplett structure
of the Hubbard bands.

The second strategy is an
asymmetric truncation scheme: instead of adding both orbital degrees 
of freedom simultaneously, the Hilbert space is truncated after 
adding each orbital individually. This also leads to a significant
reduction of the Hilbert space in the iterative diagonalization.

However, the asymmetric truncation scheme does not guarantee that orbital
symmetry is preserved during the NRG iterations. In fact, a slight
violation of orbital symmetry is observed for very small energies,
typically much lower than the Kondo temperature. It turns out
that for the DMFT-calculations presented in Sec.~\ref{sec:toHm}, both
methods give almost identical results, at least on the
scale shown in the figures.

\begin{figure}[htb]
\centerline{\includegraphics[width=0.45\textwidth,clip]{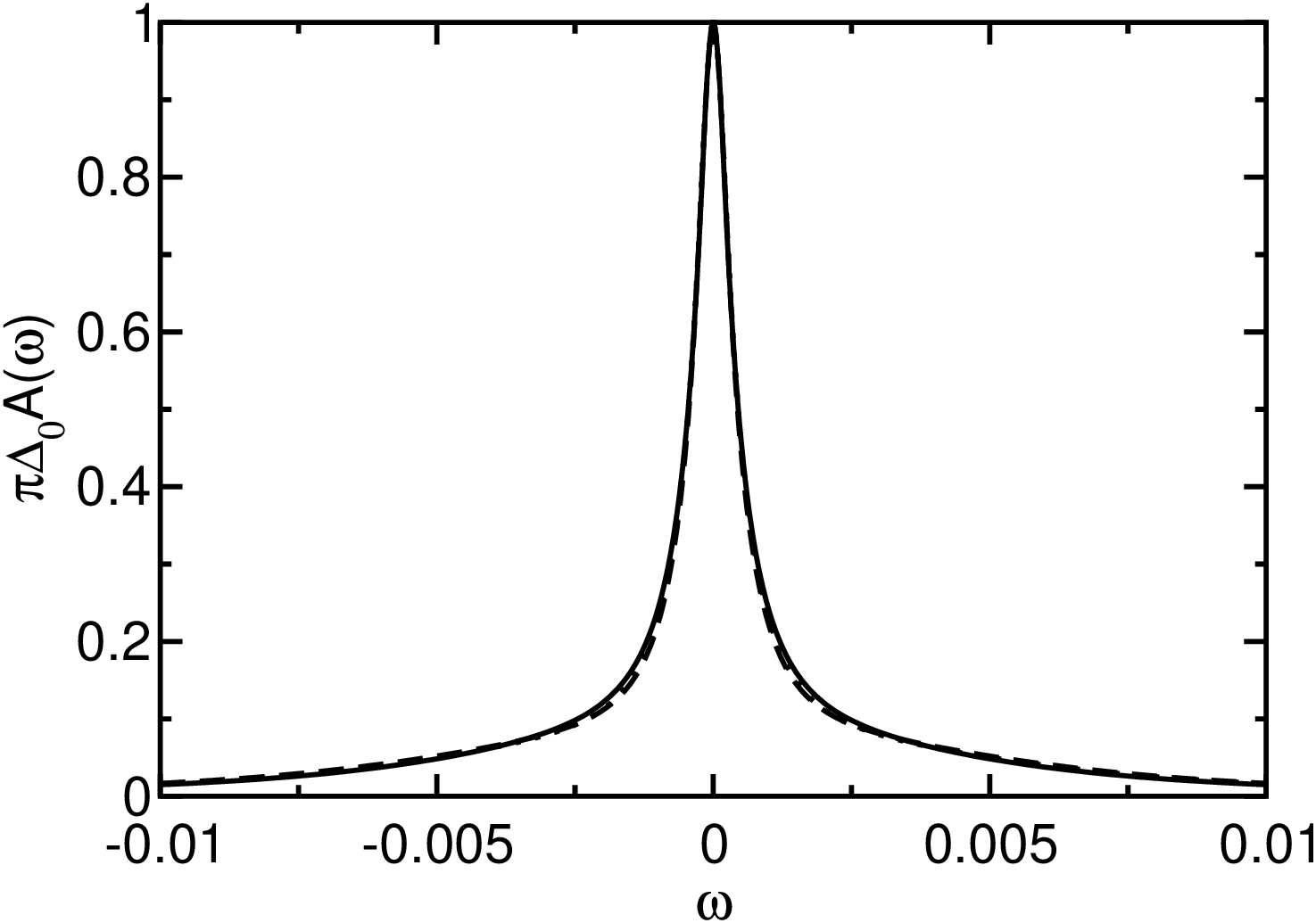}}
\caption[]{Comparison of the single-particle
dynamics for a NRG calculation (orbitally degenerate and particle-hole
symmetric single-impurity Anderson model) in the Kondo regime using 
the conventional truncation scheme (full line) and the asymmetric truncation
scheme (dashed line). Note
that particle-hole symmetry and  orbital degeneracy are preserved
even in the latter scheme.
\label{fig:compare}}
\end{figure}

Figure \ref{fig:compare} shows a comparison of the 
local single-particle density of states (DOS) for 
the impurity Anderson model eq.~(\ref{equ:tbsiam}) with twofold degeneracy,
particle-hole symmetry, and 
calculated with both schemes
(symmetric truncation with orbital quantum number 
and asymmetric truncation) in the Kondo limit.
Here, the NRG discretization parameter is $\Lambda=2.5$ 
and $3600$ states were kept in each NRG step. 
The model parameters are $U=7\Delta_0$, $J=U/100$
and $U^\prime=U-2J$. As usual, $\Delta_0=\pi
N_FV^2$
denotes the bare hybridization width ($N_F$ is the conduction electron DOS
at the Fermi energy). Obviously, the
asymmetric truncation (dashed line) leads to accurate results
for both the value of the low-energy scale and the form of the
Kondo resonance. In addition, it does not violate particle-hole symmetry 
and orbital symmetry, at least on the scale shown in 
Fig.~\ref{fig:compare}.

The aymmetric truncation scheme introduced here might be of advantage in
cases where the orbital symmetry is violated from the outset
so that the orbital quantum number cannot be used in the calculation
to reduce the matrix size.

\section{Hund's rule coupling in the single-impurity model}

Before we turn to the application of the NRG to the two-orbital Hubbard
model let us first discuss the effects of Hund's coupling for the
single impurity model (\ref{equ:tbsiam}).
For simplicity we consider a conduction band with
constant DOS, $\rho(\epsilon)=N_F$, in the interval $[-D,D]$ and choose $D=1$ as unit
of energy. The local energy $\epsilon_d$ is chosen such that the model
is particle-hole symmetric, i.e.\ $\langle n_d\rangle=2$.
As we will see, there is a profound difference between
the cases with rotationally invariant and Ising-like exchange.
The two-orbital impurity Anderson model has already been investigated
with NRG by several groups \cite{sakai_2o,cast_2o,licht_2o}.
\begin{figure}[htb]
\centerline{\includegraphics[width=0.45\textwidth,clip]{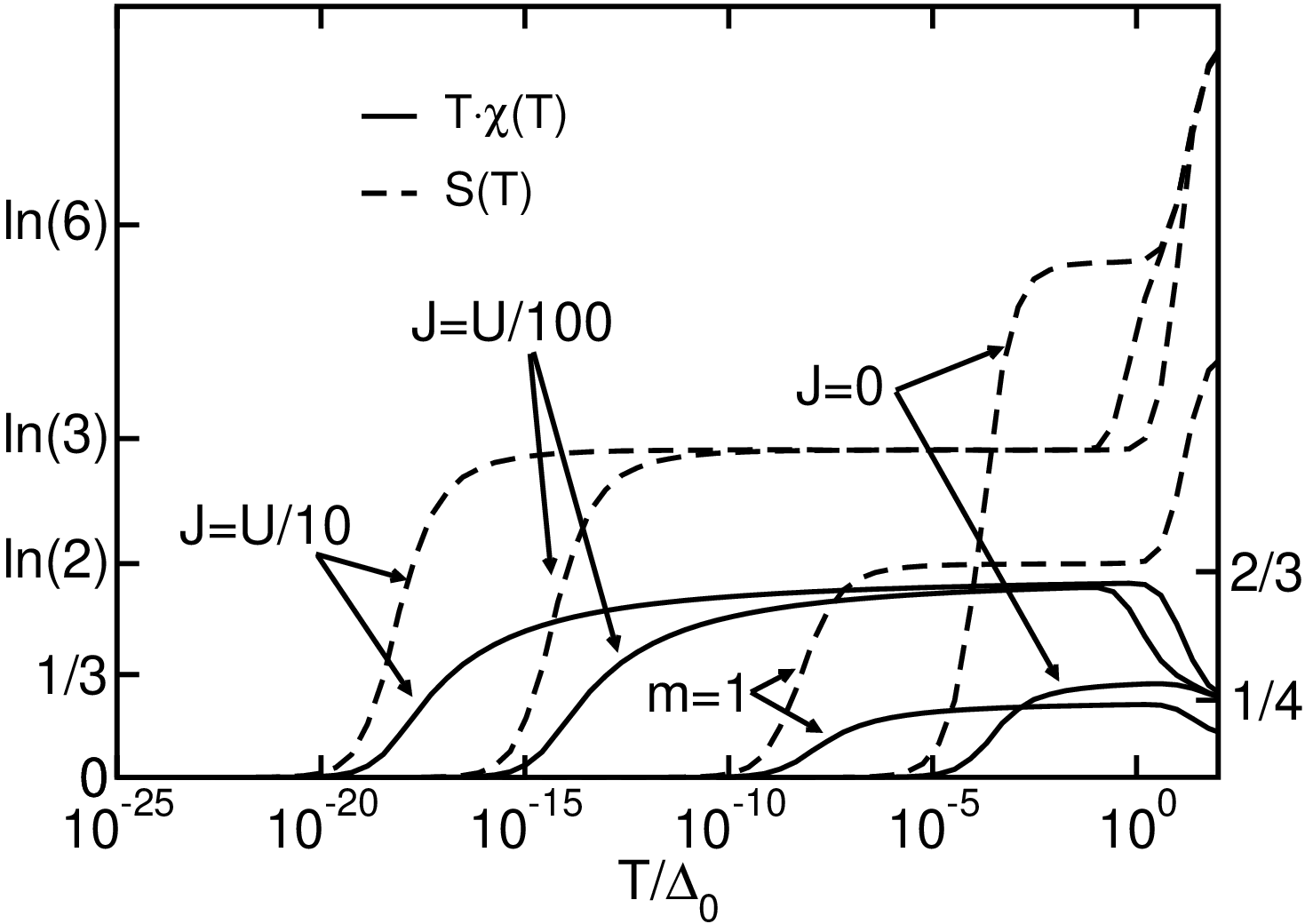}}
\caption[]{Thermodynamic properties for a particle-hole
symmetric two-orbital
impurity model with varying Hund's rule coupling $J=0$, $U/100$, and
$U/10$; dashed lines: entropy, solid lines: effective squared moments.
 The Coulomb parameter $U$ and the hybridization $\Delta_0$
were chosen such that the system is in the Kondo limit. The
inter-orbital Coulomb parameter is fixed to $U'=U-2J$ by rotational
invariance. For comparison the results of a single-orbital SIAM
are included.\label{fig:th_of_J}}
\end{figure}
Here we want to concentrate on the influence of Hund's coupling on low energy
scales and the possibility of a quantum phase transition. 
To this end we present
results for thermodynamic and dynamic properties. To improve the
accuracy of  thermodynamic properties
 we employed Oliveira's  ``Z-trick'' \cite{oliveira},
which allows to use a larger discretization $\Lambda$ (we used $\Lambda=5$)
and reduce the number of states kept ($1000$ after truncation here).

Let us begin with thermodynamic properties of the particle-hole symmetric
two-orbital single impurity model in the Kondo limit, 
that is for a hybridization
width $\Delta_0=\pi N_FV^2$  much smaller than the other bare energy
scales \cite{comment2}.
The temperature evolution of the effective squared moment\footnote{The 
adiabatic effective squared moment $T\cdot\chi_{imp}(T)$ introduced by
Wilson\cite{nrg} should not be confused with the isothermal quantity
$\langle S_z^2\rangle$. Kondo screening can be seen only in the former;
the latter goes to a finite constant value as $T\to0$.}
$\mu_{eff}^2:=T\cdot\chi_{imp}(T)$ and entropy $S(T)$ for 
$J=0$, $U/100$ and $U/10$ is shown in Fig.~\ref{fig:th_of_J}. 
The calculations were done with the rotationally invariant exchange
coupling and $U'=U-2J$, but neglecting the term breaking orbital symmetry. 
For comparison we also include results for a single-orbital SIAM
-- marked $m=1$ in Fig.~\ref{fig:th_of_J} --
with the same values of $U$ and $\Delta_0$.

For $J=0$, i.e.\ $U=U'$, we observe an intermediate ``local-moment regime'' with
entropy $S=\ln6$ and effective moment $\mu_{eff}^2=1/3$ corresponding to the six degenerate
states, two of them magnetic, in the atomic limit. As expected from general $SU(N)$
arguments \cite{hewson} this local moment is eventually Kondo screened with an 
{\em enhanced} Kondo scale $\sim\left(T_{\rm K}^{m=1}\right)^{1/m}$. As a technical
sidemark let us point out that for $J=0$, unlike finite $J$, the asymmetric
truncation does {\em not} work properly, leading to wrong results for $S$ and
$\mu_{eff}^2$ in the local moment regime. 

In the atomic limit,
any finite $J>0$ leads to a spin triplet $S=1$ as ground state with
moment $\langle S_z^2\rangle=2/3$ and
 entropy $\ln3$. Apparently, this situation
is realized for intermediate temperatures for both $J=U/100$ and $J=U/10$.
At low temperatures, this local triplet is again quenched by the conduction
electrons like for an ordinary $S=1/2$ Kondo effect.
 Obviously, the two-orbital system
has a considerably reduced low-energy
scale $T_{\rm K}$, which 
in addition decreases strongly with increasing $J$. At present we do not have a
satisfactory explanation for this observation, but believe that it is related to
the problem how spin-$1/2$ electrons screen a true $S=1$ object.

A direct consequence of this substantial reduction of the low-energy scale
for the application to the Hubbard model is that critical interactions
for an MIT are also strongly reduced for finite $J$ 
(see Sec.~\ref{sec:toHm} below).

\begin{figure}[htb]
\centerline{\includegraphics[width=0.45\textwidth,clip]{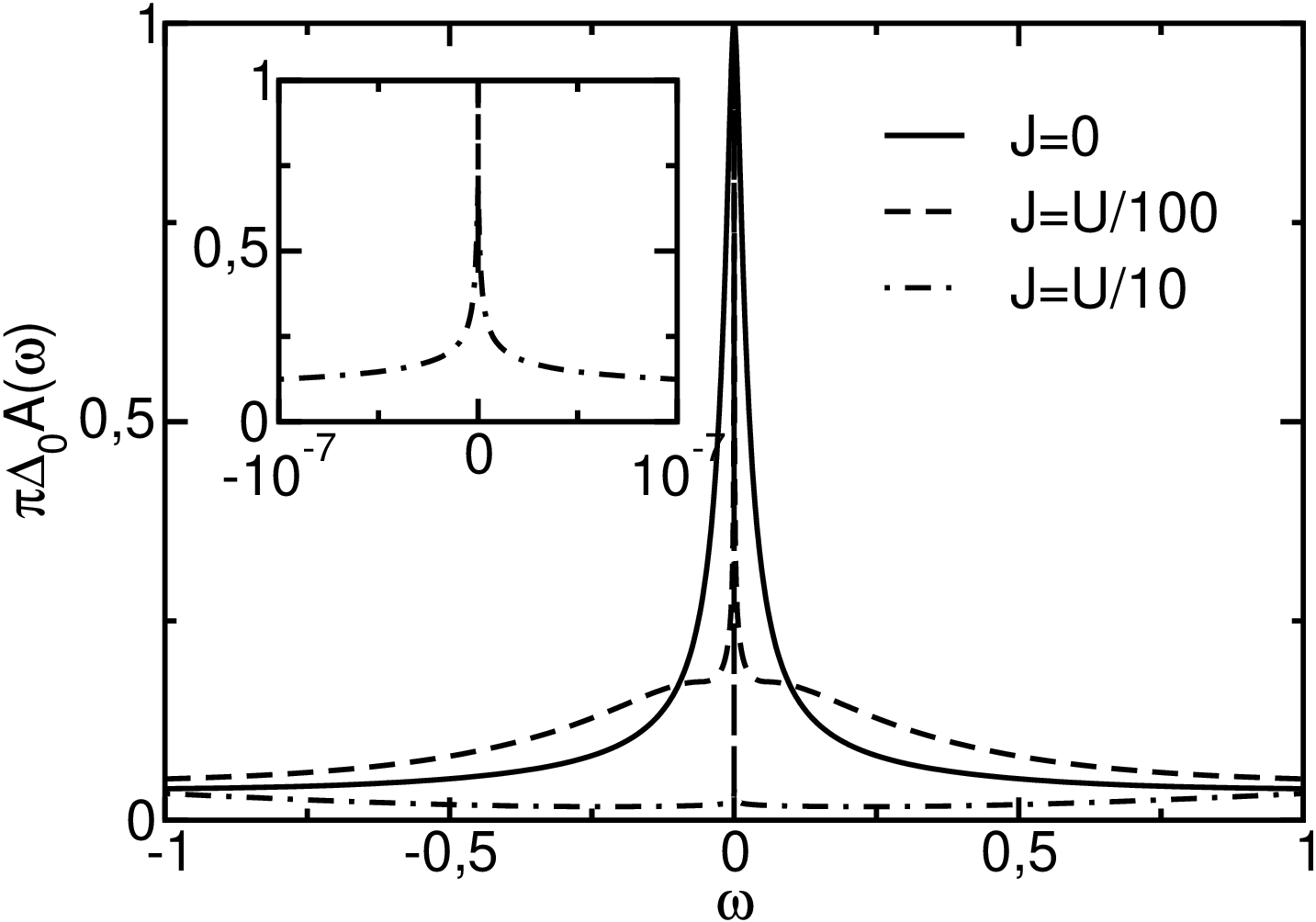}}
\caption[]{Local DOS at $T=0$ for a particle-hole
symmetric two-orbital
impurity model with Hund's rule coupling $J=0$, $U/100$ and $U/10$.
Other parameters are the same as in Fig.~\ref{fig:th_of_J}.\label{fig:dos_full_J}}
\end{figure}
The local DOS at $T=0$ for $J=U/100$ and $J=U/10$ is shown in Fig.~\ref{fig:dos_full_J}.
The calculations were done with a discretization parameter $\Lambda=2.5$
and $6400$ states kept after truncation. On the scale used in the main
panel of Fig.~\ref{fig:dos_full_J} the Kondo resonance for $J=U/10$
appears to be a vertical line, pointing to a strongly reduced Kondo 
temperature, too. In addition, new structures on the scale of $J$ appear as
shoulders in the DOS.

If one zooms into the region $[-10^{-7},10^{-7}]$ around the Fermi
energy (see inset to Fig.~\ref{fig:dos_full_J}), only the resonance for
$J=U/10$ remains visible with an energy scale well below $10^{-7}$.
This again confirms the result from the thermodynamic quantities, viz that
with increasing $J$ an exponential reduction of the Kondo temperature occurs.
It is quite obvious, that such a reduction in $T_{\rm K}$ will later manifest
itself in a corresponding reduction of the critical $U$ for the Mott-Hubbard
metal-insulator transition within the DMFT.

Let us point out that Friedel's sum rule implies as usual
the constraint $\pi\Delta_0 A(0)=1$. This constraint is fulfilled with
high precision due to the calculation of the DOS via the self-energy
according to Ref.\ \onlinecite{bulprhew}.

A completely different picture is obtained for an Ising-like exchange
interaction in model (\ref{equ:tbsiam}), which is realized by replacing
$$
J\sum_{m\ne m'}\vec S_m\cdot\vec S_{m'}
\to
J\sum_{m\ne m'}S_m^z S_{m'}^z
$$
and neglecting the term
$$
\frac{J}{2}
\sum_{m\ne m'}\sum_\sigma d_{m\sigma}^\dagger
d_{m\bar\sigma}^\dagger 
d_{m'\bar\sigma}^{\phantom{\dagger}}
d_{m'\sigma}^{\phantom{\dagger}}\;\;.
$$
In this case, the atomic ground state is doubly degenerate and consists of
\begin{figure}[htb]
\centerline{\includegraphics[width=0.45\textwidth,clip]{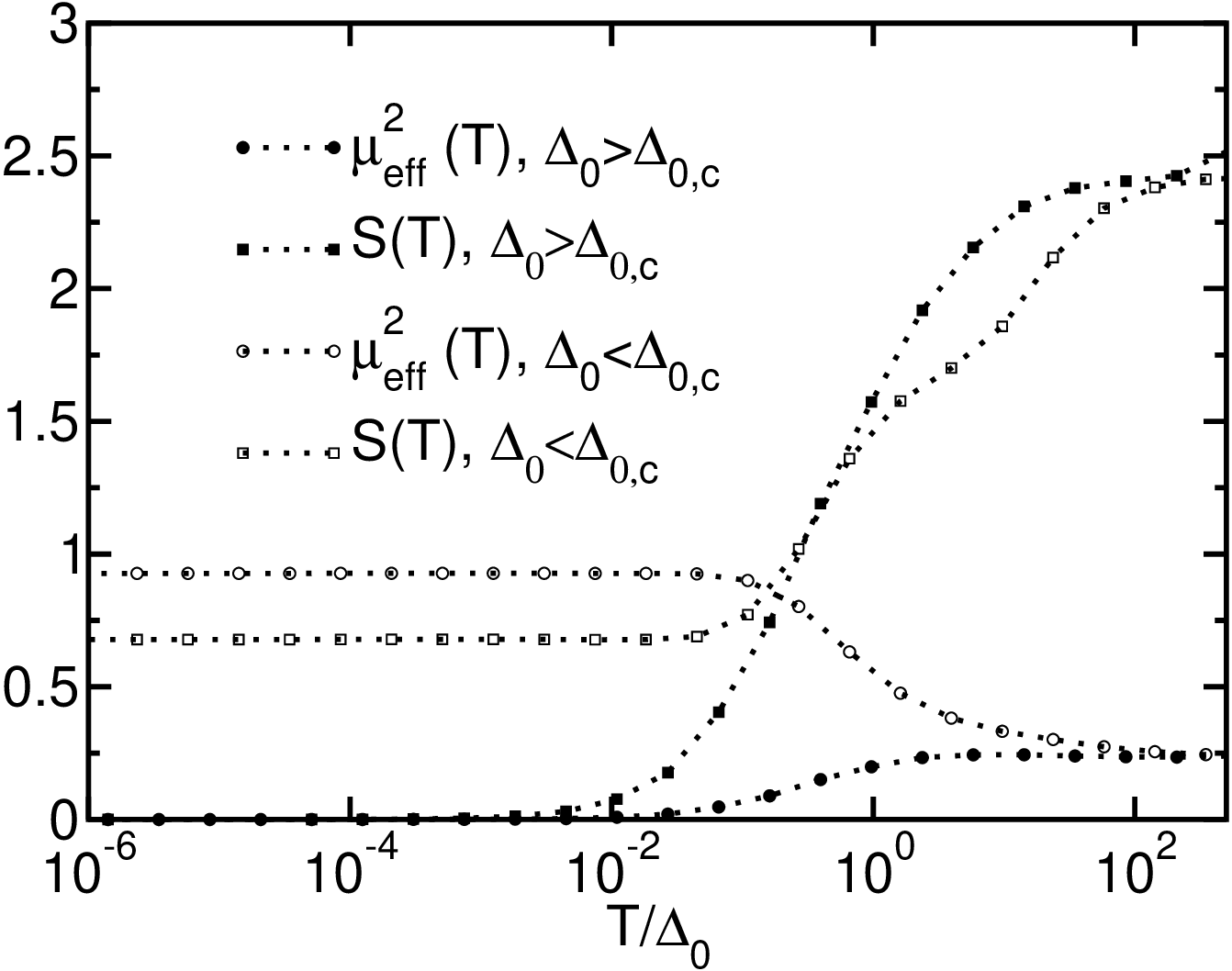}}
\caption[]{Comparison of thermodynamic properties for a particle-hole
symmetric two-orbital
impurity model with Ising-like Hund's rule coupling $J=U/100$ and two
values of the hybridization parameter $\Delta_0$. Apparently, there
exists a critical $\Delta_{0,c}$ which separates a strong coupling
fixed point ($\Delta_0>\Delta_{0,c}$) from a local moment like behaviour
($\Delta_0<\Delta_{0,c}$).\label{fig:compare_th}}
\end{figure}
the two states where two electrons with the same spin occupy different
orbitals. In contrast to the full exchange
we find that the properties change quite dramatically with
the ratio $J/\Delta_0$, where $\Delta_0$ denotes the hybridization
width. In Fig.~\ref{fig:compare_th} we compare calculations for $J=U/100$
and two different values of $\Delta_0$. For large $\Delta_0$ we find the
expected screening and corresponding formation of a Fermi liquid at low
temperatures. However, for small $\Delta_0$, this behaviour is replaced
by the formation of a state reminiscent of a local moment with entropy
$\ln2$ and effective local moment $\mu_{eff}^2\approx1$. Obviously, the neglect of the
spin-flip terms in Hund's exchange leads to a ``critical'' ratio
$J/\Delta_{0,c}$ separating strong-coupling from local moment behavior.
We interpret this feature in the following way. For the full exchange
interaction, spin-flip scattering as in the conventional $S=1/2$ case is
presumably leading to a scenario similar to the standard Kondo effect. For
Ising-like Hund coupling $J$, on the other hand, the atomic ground state
consists, as already mentioned, of the two states where two electrons with
the same spin occupy different orbitals. Quite apparently, these two
states cannot be connected by low-energy processes like spin-flips, i.e.\ the
mechanism
leading to the Kondo effect is not present here. However, if the coupling
to the band states is large enough such that the $S=1/2$ Kondo temperature
is larger than $J$, the system can screen the spins for each orbital individually
before the coupling $J$ locks the system into the states with $S_z=\pm1$,
leading to the observed strong-coupling behavior at large $\Delta_0$.

Presently, it is not clear whether the change in the impurity properties is
connected to some quantum critical behavior like in the 
pseudo-gap model \cite{pseudogapAM}.
The clarification of this question is of course of some interest in its own
right and will be discussed in detail in a forthcoming publication.

\begin{figure}[htb]
\centerline{\includegraphics[width=0.45\textwidth,clip]{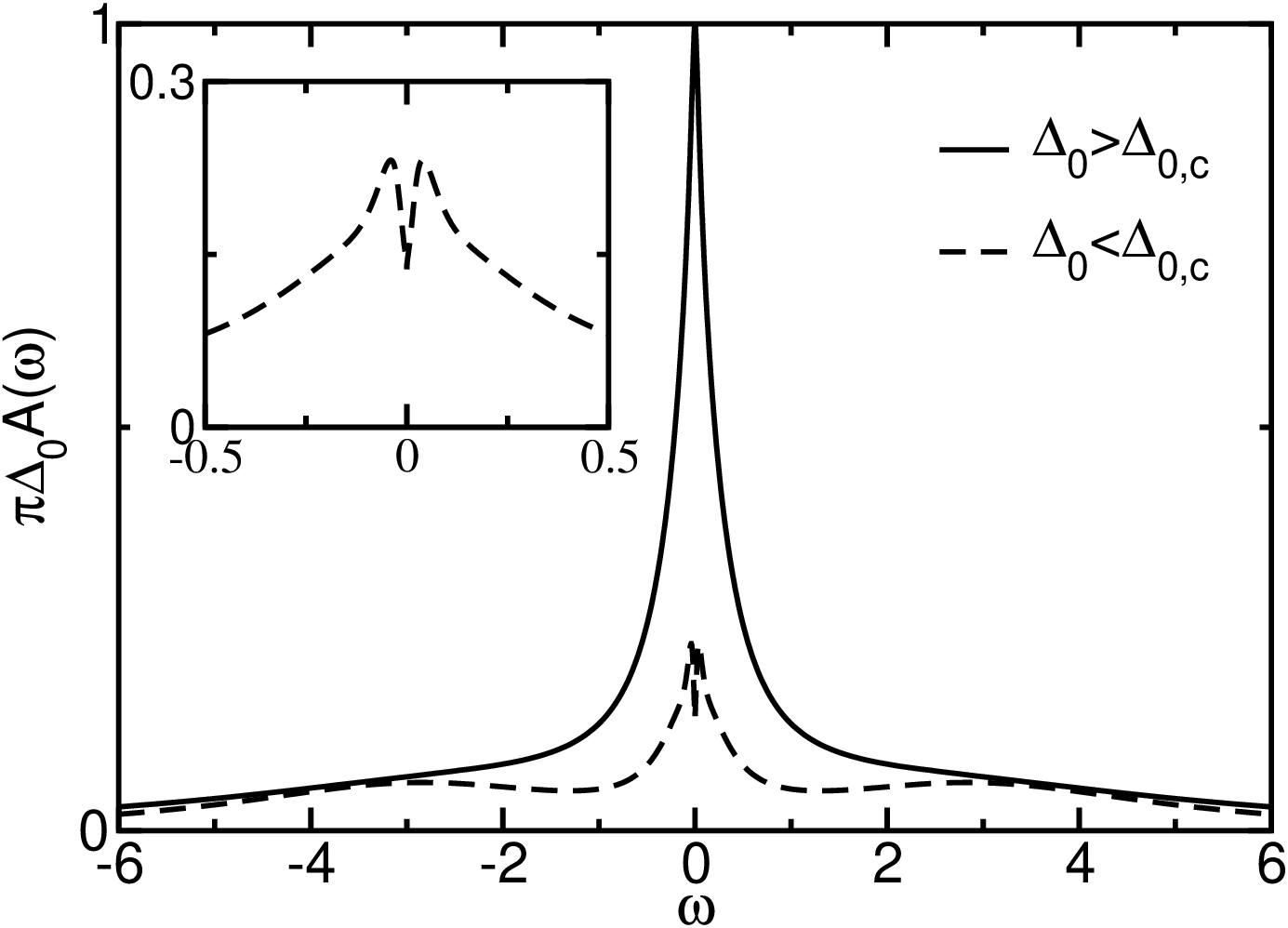}}
\caption[]{Comparison of the single-particle
dynamics for Ising-like Hund's rule coupling for hybridization strengths
above and below the critical $\Delta_{0,c}$. Parameters are the same
as in Fig.~\ref{fig:compare_th}.\label{fig:compare_dos}}
\end{figure}
The impurity DOS corresponding to the two different regimes is shown
in Fig.~\ref{fig:compare_dos}. For $\Delta_0>\Delta_{0,c}$ (full line) the
typical structure is obtained. Decreasing $\Delta_0$ below $\Delta_{0,c}$
completely changes the structure of the DOS. Instead of a Kondo peak, we
now find a structure with a pseudo-gap at the Fermi energy. It is quite
evident, that this ``criticality'' in the impurity model has profound
effects on the MIT in the DMFT calculations for the Hubbard model.

\section{MIT in the two-orbital Hubbard model}

\label{sec:toHm}

Theoretical investigations of multi-orbital Hubbard models within
DMFT have already led to a better understanding of various
issues such as the nature of the Mott-transition as a function
of orbital degeneracy 
\cite{roz97,han,Kot99,florens,kawa,ohno} 
and the structure of the spectral
function in realistic treatments within the LDA+DMFT approach
\cite{ldadmft}.
Detailed results have been obtained for the dependence of the
critical interaction strength $U_{\rm c}$ on the number of
orbitals $M$ and different integer fillings $n$. 
Numerical DMFT-QMC calculations have been performed for $M \le 3$ 
\cite{roz97,han,koga03}.
Remarkably, in the limit of large orbital degeneracy $M \to \infty$ 
an analytical treatment of the DMFT becomes possible for the MIT 
\cite{florens}. For $T=0$, a scaling $U_{\rm c} = U_{\rm c,2} \propto M$ for the 
actual transition is found
while $U_{\rm c,1} \propto \sqrt{M}$ is obtained for the critical 
interaction where the insulating solution breaks down \cite{florens}. 
This is consistent with the linear dependence for large $M$ found in 
Refs.\ \onlinecite{Lu94,Fre97,ohno} and with the square-root dependence 
reported in Ref.~\onlinecite{Koch99}. The inclusion of
the Hund's rule exchange coupling $J$ has been shown to significantly
reduce the value of $U_{\rm c}$ \cite{Koch99,han,ohno}. In particular, a qualitative
change from continuous for $J=0$ to discontinuous for any finite $J$
has been observed  in Ref.~\onlinecite{ohno} (see also the Gutzwiller results
in Ref.~\onlinecite{Bue97}).
A significant quantitative change of $U_{\rm c}$ when excluding
the spin-dependent part from the exchange coupling in 
eq.~\ref{equ:tbhubbard} has already been mentioned in 
Ref.~\onlinecite{ohno}, but detailed results
have not been published
yet.
Recently, the issue of possible orbital-selective Mott transitions
has been investigated in Refs.~\onlinecite{koga04,liebsch03}.

Let us now discuss the results from the NRG for the particle-hole
symmetric case.
To allow a direct comparison with earlier results, we use as non-interacting
DOS the semielliptic form $\rho_0(\omega)=\frac{2}{\pi}\sqrt{1-\omega^2}$
(Bethe lattice) with the same bandwidth for both orbitals. As NRG discretization
paramter we choose $\Lambda=2.5$ and keep $6400$ states after truncation. 

Except
for $J=0$, all the following results were obtained with the asymmetric truncation
scheme introduced in Sec.~II including all Coulomb interactions from
the model (\ref{equ:tbhubbard}). We have observed that the symmetric truncation
scheme (taking into account the orbital quantum number and negelcting
the last term in the interaction) leads to identical results
for spectral functions, apart from some weak redistribution of spectral weight
in the Hubbard bands due to a different atomic multiplett structure. However, the
latter effect is only barely noticable due to the broadening introduced in
calculating continous spectra from the NRG. More important, the critical values
$U_c$ for all two-band calculations within DMFT are not affected;
possible problems due to the breaking of orbital symmetry by the asymmetric truncation
scheme do not  play a role here, except again for $J=0$.

\begin{figure}[htb]
\centerline{\includegraphics[width=0.45\textwidth,clip]{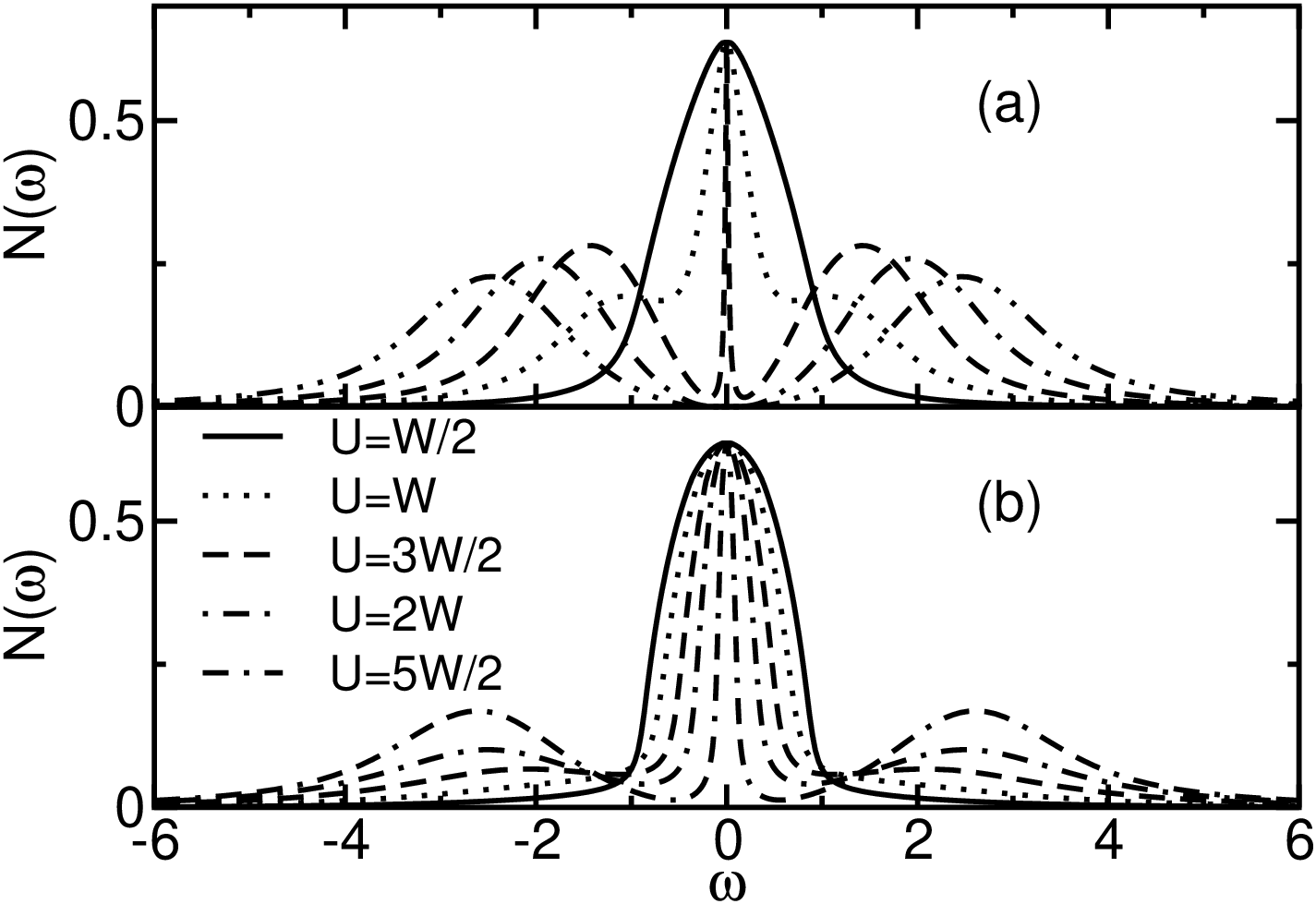}}
\caption[]{(a) Development of the DOS as function of $U$ for
(a) one-band model eq.~(\ref{equ:hubbard}) and 
(b) two-orbital model  
eq.~(\ref{equ:tbhubbard}) with $J=0$ and $U=U'$.\label{fig:2B_1}}
\end{figure}

We begin by comparing results for the one-band Hubbard model
(\ref{equ:hubbard}) (Fig.~\ref{fig:2B_1}a) and a two-orbital
Hubbard model with $J=0$ and $U=U'$ (Fig.~\ref{fig:2B_1}b).
As is well-known from earlier DMFT calculations \cite{roz97,han}, 
the critical
Coulomb parameter, $U_{c}$, for
the model with $J=0$ increases strongly with the orbital degeneracy $M$,
$U_{c}\sim M$ \cite{florens,ohno}. This is also apparent from the results in
Fig.~\ref{fig:2B_1}b.
The system stays metallic up to the largest $U$ shown. The actual MIT occurs
for a value $2.5W<U_c\lesssim3W$.

\begin{figure}[htb]
\centerline{\includegraphics[width=0.45\textwidth,clip]{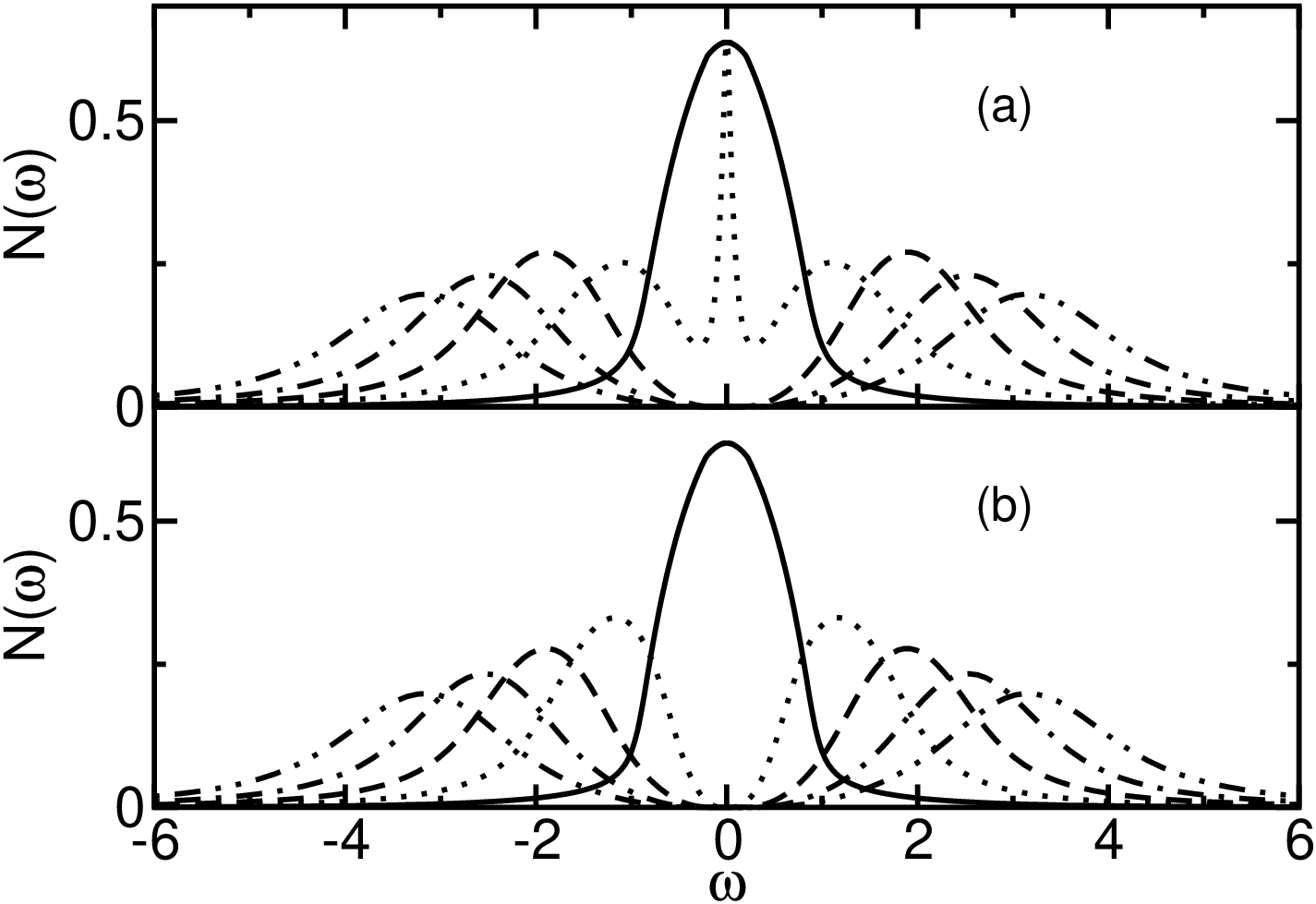}}
\caption[]{Development of the DOS as function of $U$ for the
two-orbital model with (a) full $J=U/4$ and (b) two-band model with Ising-like
$J=U/4$. In both cases $U'=U-2J$ was used. The assignment between the different
line styles and values of $U$ is the same as in Fig.~\ref{fig:2B_1}.
\label{fig:2B_2}}
\end{figure}

The results in Figs.~\ref{fig:2B_1} and \ref{fig:2B_2} 
are calculated with a fairly large value of $\Lambda=2.5$ and 
broadening parameter $b=0.8$ (see eq.~(8) in 
Ref.~\onlinecite{bulcosvol}) for both
the single-band case in Fig.~\ref{fig:2B_1}a and the two band case 
in Figs.~\ref{fig:2B_1}b and \ref{fig:2B_2}. We therefore expect that our
critical values $U_c$ for the Mott transition differ from more precise
calculations, and indeed we find them to be somewhat overestimated. 
In the single-band case, for example, our result for 
$U_c\approx 1.6W$ is slightly (10\%) larger than the well established value
$U_c=1.47 W$ \cite{bulprl}. A similar overestimation is also present for $M=2$,
where $U_c\lesssim2.5W$ is reported in literature \cite{roz97,han,kawa,florens,ohno}. 
However, the qualitative features of the transition are not altered by this
overestimation.

The influence of Hund's coupling on the development of the spectra and the
occurence of the MIT can be seen in Fig.~\ref{fig:2B_2}a,
where results for different values of $U$ and a full Hund's exchange $J=U/4$
are presented\cite{comment}.
As has been noted before \cite{kawa,ohno}, finite $J$ substantially reduces
$U_c$. Such a behavior is also seen in Fig.~\ref{fig:2B_2}a.
The critical Coulomb parameter is reduced from $U_c\approx3W$ for
$J=0$ to $U_c\approx 1.1W$.

Presently, the standard technique to solve quantum impurities with orbital degeneracy
is Quantum Monte-Carlo (QMC). However, due to the minus sign problem, one has to restrict the
Coulomb interaction to density-density type only, i.e.\ an Ising-like Hund's 
exchange. The last term in (\ref{equ:tbhubbard}) has to be
neglected completely. 
This raises the following questions: What are the consequences
of this approximation for the dynamics and in particular the MIT?

To answer this question (at least partially) we 
performed calculations with Ising-like exchange interaction
as defined in the previous section (see Ref.~\onlinecite{comment1})
The results are
shown in Fig.~\ref{fig:2B_2}b. At a first glance, the results are not
very different, except that the critical $U$ is further reduced to $U_c\approx
0.8W$. On the other hand, the results for the impurity calculation in
Figs.~\ref{fig:compare_th} and \ref{fig:compare_dos} already indicate
that the replacement of Hund's exchange by an Ising-like term has more
severe consequences than a mere quantitative change of energy scales. In the 
following we show that this approximation indeed leads to a qualitative
change in the physical properties of the Mott-Hubbard MIT.

Let us now turn to the nature of the Mott transition.
For a one-band model, it is now commonly accepted
that the transition is of second order at $T=0$ with a quasiparticle weight
that vanishes smoothly as one approaches $U_c$. There is, however, a
substantial region below $U_c$, where the insulator is metastable \cite{rmp,bulprl}. 
Previous work using the so-called linearized DMFT (L-DMFT) 
suggests that for orbitally degenerate
systems with finite $J$ this may be different \cite{ohno}. The
authors of Ref.~\onlinecite{ohno} found a first
order transition for small to intermediate $J$ signalled by a jump in the
quasiparticle weight at $U_c$. 
\begin{figure}[htb]
\centerline{\includegraphics[width=0.45\textwidth,clip]{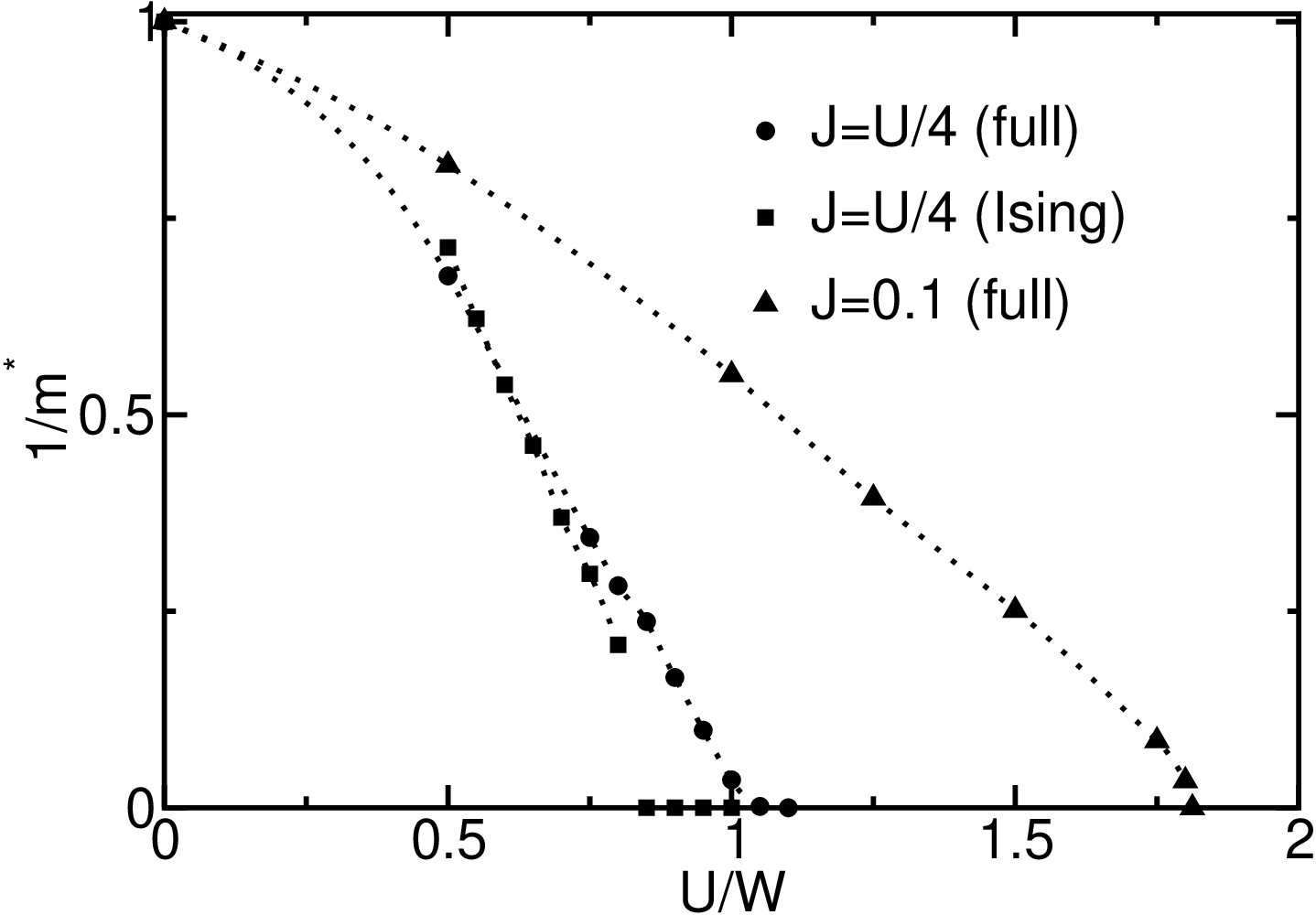}}
\caption[]{Inverse effective mass as function of $U/W$ and $J=U/4$ for the full
(circles) and Ising-like exchange interaction (squares).
The latter shows a strong jump in $1/m^\ast$ at $U_c$, leading to a first
order transition, while the former vanishes continuously.
The triangles represent a calculation with fixed $J=0.1$.
The lines are meant as
guide to the eyes.
\label{fig:mstar}}
\end{figure}
The NRG results for the inverse effective mass ($\equiv$ quasiparticle weight)
for the case $J=U/4$ are shown in Fig.~\ref{fig:mstar}. The circles were
obtained from calculations using the full interaction, while the squares
represent calculations with Ising-like interactions. Apparently,
the latter signal a strong first order transition at $U_c\approx0.8W$,
while the former lead to a continuously vanishing quasiparticle weight.
For a fixed $J=0.1$ 
(triangles in Fig.~\ref{fig:mstar}) the quasi-particle weight
near $U_c$ also shows a jump at $U_c$ as predicted by L-DMFT \cite{ohno}.
However, the magnitude of this jump comes out much smaller in our calculations.

\begin{figure}[htb]
\centerline{\includegraphics[width=0.45\textwidth,clip]{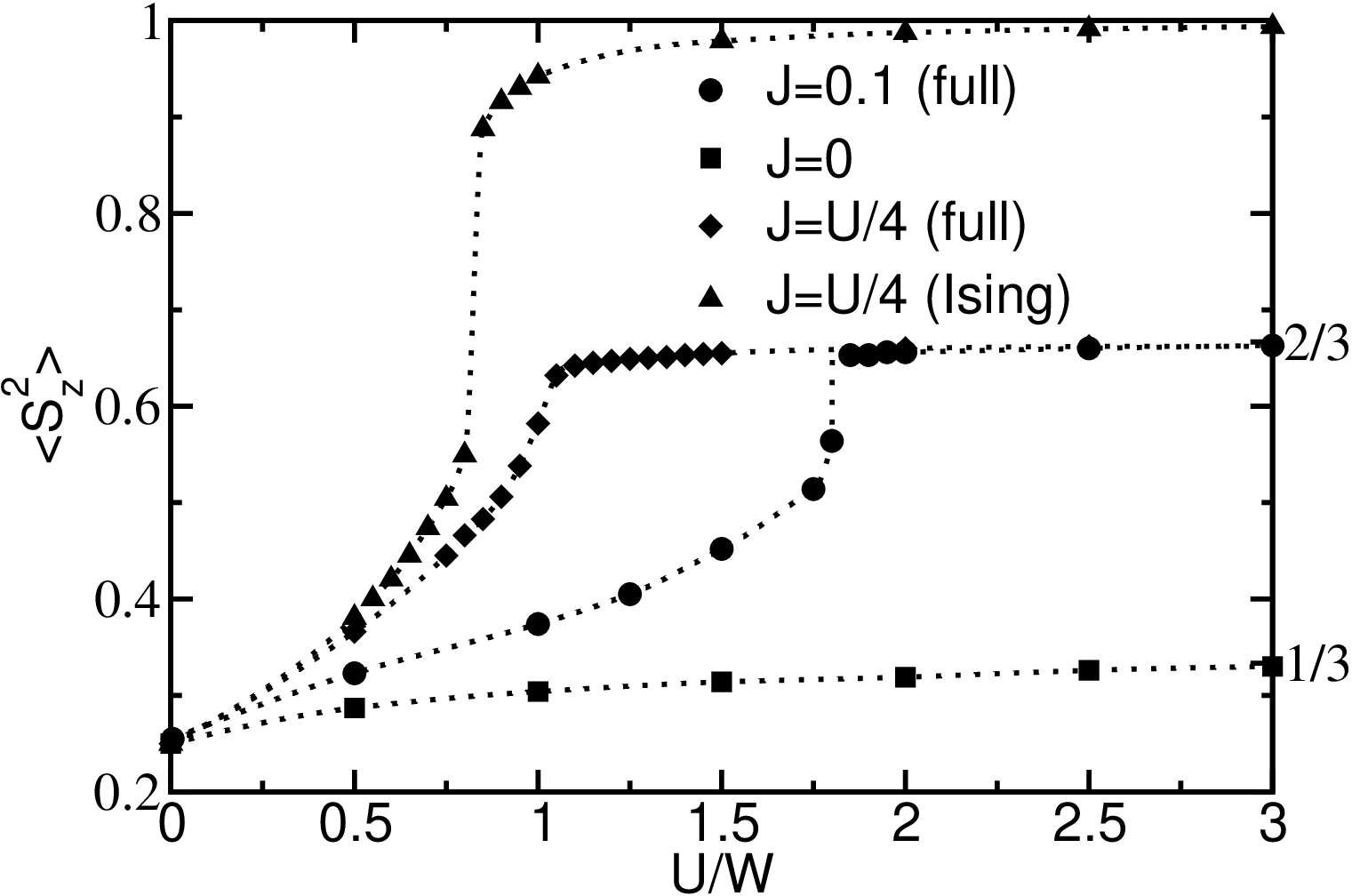}}
\caption[]{$\langle S_z^2\rangle$ for the two-band Hubbard model
and different values of $J$. The lines are meant as guide to the eye.
Note that for $J=0.1$ and Ising-like
$J=U/4$ a discontinuity in $\langle S_z^2\rangle$ occurs at the
critical $U$.\label{fig:Sz_sqr}}
\end{figure}
The differences between the different calculations ($J=0$, $J=0.1$ and $J=U/4$)
become more apparent when one looks at the local squared moment $\langle S_z^2\rangle$.
For $U\to0$ this quantity has the value $\langle S_z^2\rangle=1/4$ (for $J=0.1$
it is actually slightly larger), while deep in the Mott insulator it acquires
the atomic value enforced by Hund's coupling, i.e.\ $\langle S_z^2\rangle=1/3$ for
$J=0$, $\langle S_z^2\rangle=2/3$ for finite full $J$ and $\langle S_z^2\rangle=1$
for Ising-like $J$. This behavior is readily found in the calculated values
of $\langle S_z^2\rangle$ in Fig.~\ref{fig:Sz_sqr}. In accordance with the
results presented in Ref.~\onlinecite{ohno}, the limiting value for $\langle S_z^2\rangle$
is approached smoothly for $J=0$. The same holds for $J=U/4$, consistent
with the results for the quasi-particle weight in Fig.~\ref{fig:mstar}.
The slope, however, strongly increases when one approaches $U_c$ from below.
For constant $J=0.1$, the numerical results are not decisive, and
could be interpreted as both a
small discontinuity at $U_c$ and a continuous approach
with diverging slope.
On the other hand, a quite strong discontinuity at $U_c$ appears for an
Ising-like Hund's coupling $J=U/4$, signalling a rather strong first order
transition in this case.
The above results are in rough agreement with the L-DMFT predictions
\cite{ohno}, although there one observes a first-order transition 
also for smaller values of $J$.

The appearance of an unambiguous and rather strong first order transition
for an Ising-like exchange coupling shows that in this case the physics
underlying the Mott-Hubbard transition is very different from the one for
the rotationally invariant exchange interaction. As for the single impurity
model, we believe that a transition between individually screened orbitals
on the metallic side to a local moment regime enforced by the Ising
coupling on the insulating side occurs as soon as $J$ becomes of the
order of the Fermi liquid scale. Depending on the
details of the non-interacting DOS (its value at the Fermi level and the
band width) and the values of $U$ and $J$, this can lead in the worst case
to a serious underestimation of $U_c$ and possibly an incorrect
description of the behavior of physical quantities close to the transition.

\section{Summary and conclusions}
In this paper we presented first studies of the Mott-Hubbard transition
in a two-orbital Hubbard model within the DMFT at $T=0$ using Wilson's NRG.
In addition to a standard NRG implementation using the orbital 
quantum number, we proposed an asymmetric
truncation scheme which turns out to work rather well in both
the two-orbital single impurity Anderson model and the
two-band Hubbard model.

As a first interesting result, we observed that for the particle-hole
symmetric case a finite Hund's exchange $J>0$ leads to a
tremendous reduction in the low-energy scale $T_{\rm K}$.
This is in striking
contrast to the result for $J=0$, i.e.\ $U=U'$, where the behavior conventionally
expected for an $SU(N)$ Kondo model, viz $T_{\rm K}~\sim\sqrt[N]{T_{\rm K}^{N=1}}$, 
is found\cite{hewson}.
At present, the precise theoretical reason for this rather unexpected strong
influence of $J$ on $T_{\rm K}$ is not clear. Interestingly,
it is also rather different from the case
$\langle n_d\rangle\approx1$ where we find a mild {\em increase}
of $T_{\rm K}$ with increasing $J$.
Obviously, a detailed study of the physics of multi-orbital quantum impurity
models has to be an important future aim.

A completely different behavior occurs if one replaces the rotationally
invariant Hund exchange by an Ising-like one. In this case, the levels of the
atomic doublet with $S_z=\pm1$ enforced by the Ising coupling cannot be
connected by Schrieffer-Wolff type spin-flip processes. Thus,  the Kondo effect
can only occur for $J\alt T_{\rm K}^{m=1}$, while for larger $J$ the systems is
locked into a local moment enforced by the exchange coupling. Note that this
interpretation also implies that in the strong-coupling phase the spins
on each individual orbital will be screened separately, while for the
rotationally invariant case a full $S=1$ system must be screened.
The details of the quantum phase transition
between strong-coupling and local moment phases
have not yet been analyzed. However, in view of a possible
relevance of an Ising-anisotropy in the presence of crystal fields, a further
investigation of this problem is certainly interesting.

We also applied the NRG to the
two-orbital Hubbard model in the framework of DMFT 
to investigate the Mott-Hubbard  
metal-insulator transition at $T=0$ for the half-filled case. 
The major goal was
here to eludicate the influence of Hund's coupling on the MIT and to
investigate how the
restriction to an Ising-like exchange changes the nature of the MIT.
Our results are in general agreement with previous ones.\cite{roz97,han,kawa,florens,ohno}
In particular, for finite $J$ quantities like the effective mass or 
$\langle S_z^2\rangle$ show diverging slopes as  $U\nearrow U_c$,
possibly even discontinuities as proposed by the L-DMFT.\cite{ohno}

For an Ising-like Hund's exchange coupling the situation becomes qualitatively
different. As can be anticipated from the behavior of the 
impurity model, the MIT
is strongly first order with clear jumps in the effective mass and 
$\langle S_z^2\rangle$. Note that the former also implies a discontinuous
vanishing of the quasi-particle peak at $T=0$ as one reaches $U_c$.
Also the physics underlying this transition 
is quite different compared to the rotationally invariant case, reflecting
the lack of spin-flip scattering processes connecting the two states
$S_z=\pm1$. Thus, the metallic phase with Ising-like interaction will
be characterized by individual screening of the spins on the two orbitals rather
than a Kondo screening of a total spin $S=1$. Note that this subtlety will
most likely influence low-energy properties on the metallic side close to
$U_c$, but be less important for ``high-energy'' properties like magnetic
or orbital ordering.

It is clear, that the investigations presented here are merely a starting
point to systematically study properties of multi-orbital impurity models
or correlated lattice models within the DMFT at $T\to0$ using Wilson's NRG.
The major advantage of this method is obviously its unmatched ability
to handle exponentially small energy scales and nevertheless provide reliable
information on dynamics and thermodynamics even on high-energy scales.
Thus, at least for two-orbital models we are now in a position to
systematically study their physical properties and address questions that
are of fundamental interest for a realistic description of, for example, transition
metal oxides but require local degrees of freedom beyond a simple one-band
Hubbard model. 

\begin{acknowledgments}
We acknowledge useful conversations with
F.~Anders,
A.~Lichtenstein,
M.~Vojta,
and
D.~Vollhardt.
This work was supported by the DFG through the collaborative research center SFB 484, the
Leibniz Computer center, the Computer center of the Max-Planck-Gesellschaft
in Garching and the Norddeutsche Verbund f\"ur Hoch- und H\"ochstleistungsrechnen.

\end{acknowledgments}


\begin{thebibliography}{99}
\bibitem{comment2} The quarter-filled case is, as far as Hund's coupling
  is concerned, of little interest here. However, interesting physics can be
  found in the presence of a magnetic field (see,
  for example, Ref.\ \onlinecite{licht_2o}).
\bibitem{comment}
  We have chosen
  a fixed ration for $J/U$, because in typical experiments we expect the
  bandwidth to be the control parameter (through the variation of pressure); this means it
  is more natural to change $U/W$ and $J/W$, but keep $J/U$ fixed.
\bibitem{comment1} 
  Neglect of last term in (\ref{equ:tbhubbard})
  in fact appears to be non-critical. We 
  therefore concentrate of the spin-flip terms in Hund's coupling.
\end{thebibliography}
\end{document}